\documentclass[prb,aps]{revtex4}

\def\XXint#1#2#3{{\setbox0=\hbox{$#1{#2#3}{\int}$}
     \vcenter{\hbox{$#2#3$}}\kern-.5\wd0}}

\usepackage{graphicx}
\usepackage{bm}
\usepackage{amssymb}
\usepackage{amsmath}
\usepackage{euscript}

\begin{document}

\title{One more word on the flicker noise in manganites}

\author{Kirill~A.~Kazakov}

\affiliation{Department of Theoretical Physics,
Physics Faculty,\\
Moscow State University, $119991$, Moscow, Russian Federation}

\begin{abstract}
It is shown that the conclusion of the Comment by C.~Barone {\it et al.} [J. Appl. Phys. {\bf 115}, 116101 (2014)] on the paper [J. Appl. Phys. {\bf 113}, 094901 (2013)] is a result of several flaws including incorrect assumption that the charge carrier mobility in $La_{2/3}Sr_{1/3}MnO_3$ is independent of temperature, and misinterpretation of the quantum theory of flicker noise. An experimental evidence for strong temperature dependence of the carrier mobility in the metallic state of manganites is given, and prospects for explaining the experimental data given in the Comment are discussed.
\end{abstract}

\maketitle

According to the quantum theory\cite{kazakov1} of fundamental flicker noise, the voltage power spectrum of $1/f^{\gamma}$-noise produced by charge carriers moving in a homogeneous sample under applied voltage $V$, in the case $\gamma=1$ is given by
\begin{eqnarray}\label{mainhom}
S(f) = \frac{\varkappa V^2}{f}\,, \quad \varkappa \equiv \frac{e^3}{\pi\hbar^2
c^3}\mu T g\,,
\end{eqnarray}
\noindent where $\mu$ is the charge carrier mobility, $T$ the absolute temperature (in energy units), $e$ the elementary charge, $c$ the speed of light, $\hbar$ the Planck constant, and $g$ a geometrical factor
\begin{eqnarray}\label{gfactor}
g = \frac{1}{3\Omega}\int_{\Omega}d^3\bm{r}\left(\frac{1}{|\bm{r} - \bm{x}|} + \frac{1}{|\bm{r} - \bm{x}'|}\right),
\end{eqnarray}
\noindent $\bm{x},\bm{x}'$ denoting positions of the voltage probes, and $\Omega$ the sample volume. This expression was used\cite{kazakov3} to explain the curious results\cite{mechin2008} regarding dependence of the flicker noise intensity on the sample geometry: it turned out to be impossible to represent the noise intensity in certain manganites as a function of the sample volume, because of its essentially different dependencies on the sample width and length, but which turned out to be as predicted by Eq.~(\ref{gfactor}). However, in a recent Comment\cite{barone2014} on Ref.~2 it was suggested, on the basis of new experimental results, that the quantum theory does not correctly describe the dependence of flicker noise intensity on temperature. Namely, assuming that the charge carrier mobility is temperature-independent, the authors of the Comment inferred from Eq.~(\ref{mainhom}) that the quantum theory predicts linear dependence of flicker noise intensity on temperature, in contradiction to what they observed in manganites -- a $T$-independent $S(f)$ (see Fig.~1 of the Comment).

As will be explained below, this conclusion is a result of misinterpretation of Eq.~(\ref{mainhom}) as well as of discarding available experimental data on electric transport in manganites. This also gives me opportunity to recite the prerequisites for judicious comparison of the theory with experiment.

First of all, the assumption of temperature independence of charge carrier mobility in the studied manganite, $La_{2/3}Sr_{1/3}MnO_3,$ contradicts observations which show that below Curie temperature, compounds $La_{1-x}Sr_{x}MnO_3$ with $x>0.17$ behave as ordinary bad metals, with the carrier mobility strongly dependent on temperature.\cite{coey1999,dagotto2001,bebenin2012} This dependence is plotted in Figure~1 for $x=0.20$ (open circles) and $x=0.25$ (filled circles), and compared to that for $x=0.15.$ It is seen that in contrast to the semiconductor-type dependence in the latter case, the carrier mobility in the metallic state significantly drops as temperature increases. Though more detail on the temperature dependence of mobility is hardly found in the literature, it is known\cite{coey1999,snyder1996,bebenin2012} that the LSMO compounds tend to be more metallic as $x$ increases. In particular, decrease of their resistivity is a general trend, the residual resistivity dropping to values as small as $10^{-4}$\,$\Omega$\,cm for $x=0.3,$ comparable to those of good metals.
 
Second, Eq.~(\ref{mainhom}) does {\it not} predict linearity of $S(f)$ with respect to $T$ even when the charge carrier mobility is temperature-independent, for two reasons. One is that what is denoted $\mu$ in Eq.~(\ref{mainhom}) is not an average carrier mobility usually measured in experiments, but the highest one, {\it i.e.,} that of the most mobile carrier. In particular, the most mobile carrier is not necessarily the one dominating the electric transport, so that the temperature dependence of its mobility can be masked by the contributions of other carriers. Although this fact is clearly stated in Ref.~2 (see page 3 therein), it is ignored by the authors of the Comment. To determine $\mu$ in the case under consideration is not an easy task, as the Hall measurements needed for this are complicated by the anomalous Hall effect. In addition to that, one needs accurate data on the temperature dependence of the carrier concentrations, which is still a matter of debate in the literature. It is known that mobilities of different carriers can differ significantly, and this turns out to be the case in the material under consideration.  Namely, the electric transport in the considered compound has been thoroughly investigated at room temperature using the resistivity and Hall effect measurements, and consistency of the results has been verified on models of the electric transport in various heterojunctions involving this material.\cite{qiu2007,wang2011} These studies showed that the electron mobility at room temperature is about five times larger than that of holes. Needless to say that at different temperatures the electric transport is often dominated by different carriers.

The other implicit source of temperature dependence of the quantum flicker noise, which is often of primary importance, is the temperature dependence of the frequency exponent. In its present state, the quantum theory cannot predict the value of $\gamma,$ but it relates this value to the value of noise intensity as follows\cite{kazakov2}
\begin{eqnarray}\label{mainhom1}
S(f) = \frac{\varkappa V^2}{f^{1 +
\delta}}\,, \quad \varkappa \equiv \frac{e^3c^{\delta}}{\pi\hbar^2
c^3}\mu T g\,,
\end{eqnarray}
\noindent where now the geometrical factor reads, in the practically important case of elongated rectangular sample and small $\gamma - 1 \equiv \delta,$
\begin{eqnarray}\label{gfactorapprox1}
g = 2L^{\delta-1}/(3 \delta a^{2\delta}),
\end{eqnarray}
\noindent $a$ being the sample thickness, and $L$ its length. These formulas reveal extreme sensitivity of the flicker noise intensity to the value of the frequency exponent: collecting terms involving $\delta$ in Eqs.~(\ref{mainhom1}), (\ref{gfactorapprox1}) shows that a deviation $\delta$ in the value of this exponent is equivalent to a factor $(cL/a^2 f)^{\delta}$ in the noise magnitude. Despite smallness of $\delta,$ this factor usually takes on values up to $10$ in thin samples, because $(cL/a^2 f)$ is very large there. This result was used in Ref.~2 to estimate the calculational accuracy at $T=300\,$K, which turned out to be a factor of $5-7,$ depending on the sample size and frequency band (in the case under consideration, experimental uncertainty\cite{mechin2008} in $\gamma$ is $0.05$). But if one wants to compare noise intensity at different temperatures, it is evidently necessary to allow for the possibility of systematic changes in $\gamma$ with temperature, which even if remaining within the interval $\pm 0.05$ can easily remove the linear trend in the noise magnitude as given by Eq.~(\ref{mainhom}) with constant $\mu.$

Any of the above-mentioned flaws and omissions of the Comment is enough to cast serious doubt on its conclusion, invalidating it completely when taken together.

Still, the experimental results themselves presented in the Comment deserve further attention. In their study of the temperature dependence of flicker noise, the authors of the Comment discovered remarkable stability of the noise magnitude: the normalized Hooge parameter was found to be independent of temperature in the range $10$\,K to $300$\,K, deviations from its constant mean value in each sample being less than one percent over the entire temperature range (!) To the best of my knowledge, such extraordinary constancy has never been observed previously, neither in manganites nor in other materials. Even in good metals where the law $\mu\sim 1/T$ holds fairly well down to $T\approx 50$\,K, experiments show at least $5\%-10\%$ deviations of the noise magnitude from its constant mean (see, {\it e.g.}, Ref.~12). These new experimental findings, therefore, present a real difficulty for the flicker noise theory. In fact, while it is conceivable that all the physical parameters characterizing electric transport and noise in manganites could be determined within the one percent accuracy, it is highly unlikely that they would combine into a $T$-independent constant upon substitution into the formula (\ref{mainhom1}). This is actually difficult to expect of any known physical mechanism of flicker noise (charge carrier trapping-detrapping, defect motion, {\it etc.}), the more so as the noise magnitude in the same samples does depend on temperature above $300$\,K (see Fig.~7 of Ref.~3).

\begin{figure}
\includegraphics[width=0.4\textwidth]{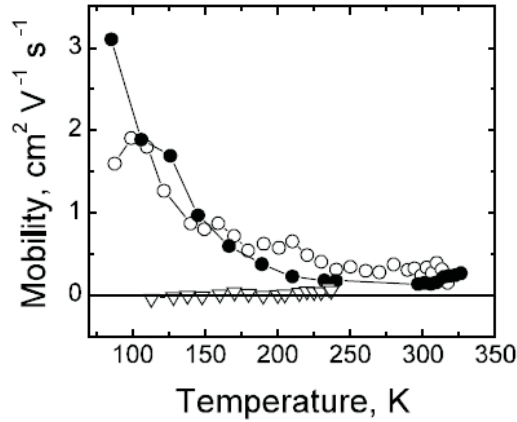}
\caption{Hall mobility versus temperature in $La_{1-x}Sr_{x}MnO_3$ for $x=0.15$ (triangles), $x=0.20$ (open circles), and $x=0.25$ (filled circles). {\it Source}: Fig.~2 of Ref.~7. Reprinted with permission from Journal of Magnetism and Magnetic Materials {\bf 324}, 3593 (2012). Copyright 2012 Elsevier.}
\end{figure}

\pagebreak

\end{document}